\documentclass[preprint,pre,sort&compress]{revtex4}
\input epsf

\def \nn {\nonumber}


\begin{document}
\title{Structural changes of diblock copolymer melts due to an
external electric field: \\
a self-consistent field theory study}
\author{Chin-Yet Lin}
\author{M. Schick}
\affiliation{Department of Physics, University of Washington,
Seattle, Washington 98195-1560}
\author{David Andelman}
\affiliation{School of Physics and Astronomy, Raymond and Beverly
Faculty of Exact Sciences, Tel Aviv University, Ramat Aviv 69978,
Tel Aviv, Israel}
\date{February 25,2005}

\begin{abstract}

We study the phase behavior of diblock copolymers in presence of
an external electric field. We employ self-consistent field theory
and treat the relevant Maxwell equation as an additional
self-consistent equation. Because we do not treat the electric
field perturbatively, we can examine its effects even when its
magnitude is large. The electric field couples to the system's
morphology only through the difference between the dielectric
constants of the two blocks. We find that an external field aligns
a body-centered cubic phase along the (111) direction, reducing
its symmetry group to $R{\bar 3}m$. Transitions between this phase
and the disordered or hexagonal phases can occur for external
electric fields ranging from a minimum to a maximum value beyond
which the $R{\bar 3m}$ phase disappears completely. This
electric-field range depends on diblock architecture and
temperature. We present several cuts through the phase diagram in
the space of temperature, architecture and applied field,
including one applicable to a system recently studied.

\end{abstract}

\maketitle

\section{Introduction}

Because block copolymers readily self-assemble into various
ordered arrays, they have been avidly studied for technological
applications such as  high density porous materials,
nano-lithographic templates, photonic band gap materials
\cite{Park97,NPF99,LCM96}, and well-ordered arrays of  metal
nano-wires \cite{Park97,UNA00}. One practical difficulty to their
use in some applications is that the ordered phase is not created
in one single crystal, but rather in domains of differing
orientation. One means of aligning the domains is to apply an
external electric field. It has been shown
\cite{LCM96,OII00,Boker02,Boker03,Tsori02} that applying an
electric potential, on the order of a few to a few dozen volts across
electrodes separated by several micrometers, can effectively orient
domains of lamellar or cylindrical morphology normal to the
surfaces of thin films. The basis of this orientation effect is
simple. In order to reduce accumulation of polarization charge,
the system lowers its free energy by aligning cylinders or
lamellae so that their long axis is parallel to the applied field.

Recently, related experiments on diblock copolymers \cite{Xu04}
have been performed where external fields have been applied to
bring about a phase transition from a phase of spheres to one of
cylinders. In the phase of spheres, it is not possible to
eliminate the accumulation of polarization charge so that its free
energy increases in an external field with respect to a
cylindrical phase, and a phase transition can be induced. This
change in phase due to the application of an electric field was
considered by Tsori et al. \cite{Tsori03}, and by Xu et al
\cite{Xu04}.

The effects of an external field on an ordered array of
inhomogeneous dielectric material is of great interest. First, the
problem is inherently self-consistent simply because the material
is a dielectric; {\em i.e.} the electric field at a given point
depends upon the polarization at that point which, in turn,
depends upon the local electric field. In addition, in the problem
of interest here, the local dielectric constant is inhomogeneous.
It depends upon the morphology of the ordered phase, which itself
depends upon the local electric field \cite{Amundson94}.

In previous calculations for diblock copolymers
\cite{Tsori03,Xu04}, this self-consistent circle has been broken
by assuming that the two block are only weakly segregated,
resulting in a small amplitude of the spatial variation of the
relative concentration of the two blocks.  In this case it follows
from the vanishing of the divergence of the electric displacement
that the amplitude of the spatially varying electric field is also
small so that the electrostatic Maxwell equation can be solved
perturbatively for the electric field as a function of the order
parameter. This procedure was carried out to quadratic order in
the field by Amundson et al.\cite {Amundson93}.  It is appropriate
in the weak-segregation limit, and should be adequate for
determining the general phase behavior in weak external fields.
However, since experiments are often not in the weak segregation
limit, and the effect of electric fields has hardly been explored,
further study is clearly called for.

In recent years, thermodynamic properties of block copolymer
systems have been treated successfully by the {\em full}
self-consistent field (SCF) theory, to which weak- and
strong-segregation theories are approximations \cite{Matsen94}.
Given the self-consistent nature of an inhomogeneous dielectric in
an external electric field, it seems natural to apply the full SCF
theory to this problem as well. That is what we do in this paper.
We solve exactly the full set of SCF equations {\em and} the
appropriate Maxwell equations under the assumption of a simple
constitutive relation between the local dielectric properties and
the local volume fractions. In particular, we consider the
evolution of the bulk phase diagram of diblock copolymers in an
applied electric field, and focus upon its effect on reducing the
region of the phase diagram occupied by the body-centered cubic
(bcc) structure, (space group $Im{\bar 3}m$). Evolution of the
gyroid structure (space group $Ia{\bar 3}d$), whose region in the
phase diagram also decreases due to the application of a field, is
not considered.

We calculate the strength of an external field needed to bring
about a phase transition from the (distorted) spherical phase to
the disordered phase and to the cylindrical phase. For the transition to
the latter phase we find two distinct behaviors depending upon the
architecture of the diblock, as measured by the parameter $f_A$
introduced below. The first is brought about if a transition
from the spherical phase to the cylindrical phase can be induced
in the absence of an external electric field
simply
by reducing the temperature in the realm of interest.
If so, the same must also be true for very small
fields. As a consequence, one can always find
a temperature in that realm at which an arbitrarily small field will induce a
transition from the spherical to the cylindrical phase.
The other behavior occurs if
the spherical phase is the most stable one in the absence of an external
field for temperatures in the realm
of interest. In that case, a {\em non-zero}
external field is required to induce a transition from it to the
cylindrical one at any temperature in this realm.
In either case, we
find that for a given architecture, there is a maximum value of applied field
beyond which the spherical phase is no longer the most stable one for
any temperature.

In the following section, we set up the general formalism. In
section III, we discuss its application to the phase of
(distorted) spheres, and compare the results of the full
self-consistent calculation with those obtained from an expansion
of the free energy in the electric field to order $E^2$. Such an
expansion does not indicate the optimal direction in which
the field aligns the cubic phase, whereas the full calculation
shows that alignment along the (111) direction is
favored over a (100) orientation. There is a concomitant reduction
of the symmetry of the phase from $Im\bar{3}m$ (bcc phase) to
$R\bar{3}m$ (distorted spherical phase). Various cuts through the
phase diagram are also presented. We conclude with a brief summary
and comparison with recent experiments.

\section{GENERAL FORMALISM}

We consider a melt of $n$ A-B diblock copolymer chains, each of
polymerization index $N=N_A+N_B$. If the specific volumes of the A
and B monomers are $v_A$ and $v_B$, respectively, the volume per
chain is $v_p=N_Av_A+N_Bv_B$. For an incompressible melt of A-B
chains, the volume fraction of the A monomers is
$N_Av_A/(N_Av_A+N_Bv_B)$, and the total system volume is
$\Omega=nv_p$. We assume the monomer volumes to be identical,
$v_A=v_B$, so that the volume fraction of the A-monomers
is equal to the mole fraction of the A-monomers, $f_A=
N_A/N$. We also assume that the Kuhn lengths of the $A$ and $B$
components are identical, a length denoted $a$.

In the absence of an external field, the application of SCF theory
\cite{Schmid98} leads to a free energy ${\cal F}$ which is a
functional of unknown fields $W_A$, $W_B$, and $\Xi$, and a
function of temperature $T$
\begin{eqnarray}
\frac{{\cal F}(W_A,W_B,\Xi;T)}{nk_B T} &\equiv&
-\ln\mathcal{Q}[W_A,W_B] +\frac{1}{\Omega}\int d\mathbf{r} \left\{
\chi N\Phi_A\Phi_B -
W_A\Phi_A - W_B\Phi_B ~\right. \nn \\
&& \left.  \qquad - ~\Xi(1-\Phi_A-\Phi_B)\right\}~,
\end{eqnarray}
where $k_B$ is the Boltzmann constant, $\Phi_A({\bf r})$ and
$\Phi_B({\bf r})$ are the local volume fractions of A and B
monomers. The dependence on $T$ comes from the usual Flory
interaction parameter, $\chi$, which to a good approximation
is inversely proportional to the temperature, $\chi N=b/T$ with $b$
a constant. The function $\mathcal{Q}[W_A,W_B]$ is the partition
function of a single polymer chain
subject to the fields $W_A({\bf r})$ and $W_B({\bf r})$, as is
given below. The field $\Xi({\bf r})$ is a Lagrange multiplier
that enforces locally the incompressibility constraint,
$\Phi_A({\bf r})+\Phi_B({\bf r})=1$. The three unknown fields are
determined by requiring that the free-energy functional be
extremized with respect to their variation at constant $T$.

The fields $W_A$ and $W_B$ appear in the single-chain partition
function of the flexible diblock copolymer, ${\mathcal
Q}[W_A,W_B]=\int d{\bf r}q({\bf r},1)/c$, where $q({\bf r},s)$
satisfies the modified diffusion equation
\begin{equation}
\frac{\partial q}{\partial s} = \frac{1}{6}Na^2\nabla^2q-W_A({\bf
r})q,\qquad\ {\rm if~~}\ 0\leq s \leq f_A ~,
\end{equation}
and
\begin{equation}
\frac{\partial q}{\partial s} =
\frac{1}{6}Na^2\nabla^2q-W_B({\bf r})q, \qquad \ {\rm if~~}\ f_A\ <\ s\
\leq 1 ~,
\end{equation}
with the initial condition $q({\bf r},0)=1$, and $c$ is a volume of no
consequence here.

The addition of a local electric field ${\bf E}({\bf r})$ in the
derivation of the free energy ${\cal F}$ is straightforward. In an
ensemble for which an external electric potential is held fixed
\cite{Landau84}, the above free energy simply becomes
\begin{eqnarray}
\label{intfreeenergy} \frac{{\cal F}(W_A,W_B,\Xi;T,{\bf E})}{nk_B
T} &=& -\ln\mathcal{Q}[W_A,W_B] - \frac{\epsilon_0v_p}{k_B T}\int
\frac{d\mathbf{r}}{2\Omega}\kappa(\mathbf{r})|\mathbf{E(\mathbf{r})}|^2\nn \\
&+&\frac{1}{\Omega}\int d\mathbf{r} \left\{ \chi N\Phi_A\Phi_B -
W_A\Phi_A - W_B\Phi_B - \Xi(1-\Phi_A-\Phi_B)\right\} ~,
\end{eqnarray}
where $\epsilon_0$ is the vacuum permittivity, and $\kappa({\bf
r})$ is the local dielectric constant. A constitutive relation
between $\kappa({\bf r})$ and the volume fractions of A and B
monomers must be specified. We choose a linear interpolation
relation

\begin{equation}
\kappa({\bf r})=\kappa_A\Phi_A({\bf r})+ \kappa_B\Phi_B({\bf r})
~,
\end{equation}
where $\kappa_A$ and $\kappa_B$ are the dielectric constants of
pure A and B homopolymer  phases, respectively. This choice is
clearly correct in the limiting cases of the pure systems, and
in the weak-segregation limit.
It also has the virtue of simplicity and should capture the correct physics.

From the above it can be seen that a convenient scale for the
strength of the electric field is
\begin{equation}
{\cal E}\equiv \left(\frac{k_BT}{\epsilon_0 v_p}\right)^{1/2}~.
\end{equation}
The magnitude of this electric field unit at typical experimental
temperatures, $T\simeq 430$\,K, and for typical volume per polymer
chain, $v_p\simeq 100$\,nm$^3$, is ${\cal E}\simeq 82$\,V/$\mu$m.
We shall denote the dimensionless electric field rescaled in this
unit as ${\hat{\bf E}}\equiv{\bf E}/{\cal E}$. Similarly a
dimensionless displacement field, ${\hat{\bf D}}$, is conveniently
defined by ${\hat{\bf D}}\equiv{\bf D}/\epsilon_0{\cal E}.$

The requirement that the free energy functional be an extremum
with respect to variation of $W_A$, $W_B$, $\Xi$, and of the
volume fractions $\Phi_A$ and $\Phi_B$ at constant temperature, or
$\chi N$, and fixed electric field $\hat{\mathbf{E}}$, leads to
the following set of SCF equations:

\begin{eqnarray}
\label{sceq1}
w_A&=&\chi
N\phi_B+\xi-\frac{1}{2} \kappa_A|\hat{\mathbf{E}}|^2 ~,\\
w_B&=&\chi
N\phi_A+\xi-\frac{1}{2} \kappa_B|\hat{\mathbf{E}}|^2 ~,\\
\phi_A+\phi_B & =& 1 ~,\\
\phi_A&=&-\frac{\Omega}{\mathcal{Q}}\frac{\delta\mathcal{Q}}{\delta w_A} ~,\\
\label{sceq5}
\phi_B&=&-\frac{\Omega}{\mathcal{Q}}\frac{\delta\mathcal{Q}}{\delta
w_B} ~.
\end{eqnarray}
The values of $W_A$, $W_B$, $\Xi$, $\Phi_A$ and $\Phi_B$, which
satisfy these equations are denoted by lower case letters, $w_A$,
$w_B$, $\xi$, $\phi_A$, $\phi_B$, respectively. The free
energy within the SCF approximation, $F_{\rm scf}$ is obtained by
substitution of these values into the free energy of
Eq.~(\ref{intfreeenergy}),
\begin{equation}
\label{mffreeenergy} F_{\rm scf}(T,{\bf E})={\cal
F}(w_A,w_B,\xi;T,{\bf E}) ~,
\end {equation}
or
\begin{equation}
\frac{F_{\rm scf}}{nk_B T} = -\ln\mathcal{Q}[w_A,w_B] -
\frac{1}{\Omega}\int d\mathbf{r} [ \chi
N\phi_A(\mathbf{r})\phi_B(\mathbf{r}) +\xi(\mathbf{r})]
~.\label{mffreeenergy1}
\end{equation}

In addition to these equations, there are also the Maxwell equations which
the electrostatic field must satisfy in absence of free charges:
\begin{eqnarray}
\nabla\times{\hat{\bf E}}&=&0 ~,\\
\nabla\cdot{\hat\mathbf{D}}(\mathbf{r})&
\equiv&\nabla\cdot(\epsilon_0\kappa(\mathbf{r})
{\hat\mathbf{E}}(\mathbf{r}))=0 \label{maxwell} ~.
\end{eqnarray}
As usual, we guarantee that the first of these equations is
satisfied by introducing the electric potential $\hat{V}({\bf
r})$,

\begin{equation}
{\hat{\bf E}}({\bf r})=-\nabla \hat{V}({\bf r})=-\nabla {V}({\bf r})/{\cal
E} ~.
\end{equation}

Since we will consider, in addition to the disordered phase,
spatially-periodic ones, it is convenient to write all functions
of position in terms of their values averaged over a {\em unit
cell}
\begin{equation}
C_0\equiv \langle C\rangle = \frac{\int_{\rm unit\ cell}C({\bf
r})\ d{\bf r}}{\int_{\rm unit\ cell}\ d{\bf r}} ~,
\end{equation}
and their deviations from those average values
\begin{equation}
\delta C({\bf r})\equiv C({\bf r})-C_0 ~.
\end{equation}
The average values of several quantities of interest are
\begin{eqnarray}
\phi_{A,0}&=&f_A ~,\nonumber\\
\phi_{B,0}&=&1-f_A ~,\nonumber\\
\kappa_0&=&\kappa_Af_A+\kappa_B(1-f_A) ~,\nonumber\\
w_{A,0}&=&\chi N(1-f_A)-\frac{1}{2}\kappa_A|{\hat{\bf E}}_0|^2 ~,\nonumber\\
w_{B,0}&=&\chi N f_A-\frac{1}{2}\kappa_B|{\hat{\bf E}}_0|^2 ~,
\end{eqnarray}
where the value of $\xi_0$ has been arbitrarily set to zero, and
$\hat{\bf E}_0$ is the value of the local electric field averaged
over a unit cell. To determine this without knowing the full
spatially dependent electric field ${\bf E}({\bf r})$, we reason
as follows. Assume that the external field is produced by planar
electrodes which are separated by a distance $d$ and subject to a
voltage difference ${V}_{12}$. In the gap, and along the $z$-axis
perpendicular to the electrodes, the field is $E_{\rm
ext}=-V_{12}/L$. Given that the dielectric fills the space between
the plates, and that the voltage $V_{12}$ is held fixed as the
dielectric is inserted, it follows that $\int E_z\,dz=E_{\rm
ext}L$, and that the average value of $E_z$ is $E_{\rm ext}$. We
make a reasonable assumption that the free energy of the system is
minimized when an axis of symmetry of one of the ordered
structures coincides with the $z$-axis. In this case $E_0=\int_0^L
E_z dz/L=E_{\rm ext}$.
Hence in rescaled units
\begin{equation}
{\hat{\bf
E}}_0=\left(\frac{\epsilon_0v_p}{k_BT}\right)^{1/2}E_{\rm ext}\
{\hat{\bf z}} ~,
\end{equation}
and
\begin{equation}
\delta{\hat{\bf E}}({\bf r})= \hat{\bf E}({\bf r}) - \hat{\bf
E}_0\equiv -\nabla\delta {\hat V}({\bf r}) ~.
\end{equation}

Utilizing these average values, we can rewrite the free energy in
the SCF approximation,
Eqs.~(\ref{mffreeenergy})-(\ref{mffreeenergy1}), in the form
\begin{equation}
\frac{F_{\rm scf}}{nk_BT}=
-\ln\left\{\frac{{\cal Q}[w_A,w_B]}{{\cal Q}[w_{A,0},w_{B,0}]}\right\}+
\chi Nf_A(1-f_A)-\frac{ 1}{2}\kappa_0{\hat E}_0^2-\frac{\chi
N}{\Omega}\int \delta\phi_A({\bf r})\delta\phi_B({\bf r})d{\bf r}
~,
\end{equation}
where, from the incompressibility condition, $\delta\phi_A({\bf
r})=-\delta\phi_B({\bf r})$. Note that the electric-field
contribution $-\kappa_0{\hat E}_0^2/2$ is common to all phases.
For the lamellar and hexagonal phases in the lowest energy
orientation, this is the {\em only} contribution to the free
energy from the electric field.
It
can conveniently be absorbed in a redefinition of the free energy
\begin{equation}
f_n({\hat E}_0)\equiv F_{\rm scf}/nk_BT+\frac{1}{2}\kappa_0{\hat
E}_0^2 ~. \label{fn}
 \end{equation}

The advantage of separating out the average values is that the
only remaining  Maxwell equation, Eq.~(\ref{maxwell}), can be
written as an {\em inhomogeneous}  equation for the potential
$\delta \hat{V}({\bf r})$,
\begin{eqnarray}
\label{maxeq} & &\nabla\delta \hat{V}({\bf
r})\cdot\nabla[\kappa_A\delta \phi_A({\bf r})
+\kappa_B\delta\phi_B({\bf r})] +[\kappa_A(f_A+\delta\phi_A({\bf
r}))+\kappa_B(1-f_A+\delta\phi_B({\bf r}))]\nabla^2\delta
\hat{V}({\bf
r})\nonumber\\
& &={\hat E}_0\frac{\partial}{\partial
z}[\kappa_A\delta\phi_A({\bf r})+\kappa_B\delta\phi_B({\bf r})] ~.
\end{eqnarray}
This, with the three
remaining self-consistent equations,
\begin{eqnarray}
\delta w_A({\bf r})&=&\chi N\delta\phi_B({\bf r})+\delta\xi({\bf
r})+\frac{1}{2}\kappa_A[2{\hat{\bf E}}_0\cdot\nabla\delta
\hat{V}({\bf
r})-(\nabla\delta \hat{V}({\bf r}))^2]~,\\
\delta w_B({\bf r})&=&\chi N\delta\phi_A({\bf r})+\delta\xi({\bf
r})+\frac{1}{2}\kappa_B[2{\hat{\bf E}}_0\cdot\nabla\delta
\hat{V}({\bf
r})-(\nabla\delta \hat{V}({\bf r}))^2]~,\\
\label{inc} \delta\phi_A({\bf r})&+&\delta\phi_B({\bf r})=0 ~,
\end{eqnarray}
constitute the four self-consistent equations which determine the
four functions $\delta w_A({\bf r})$, $\delta w_B({\bf r})$,
$\delta \xi({\bf r})$, and $\delta \hat{V}({\bf r})$.

We note that with our choice of constant external field applied
along the $z$ direction,  the Maxwell equation, Eq.~(\ref{maxeq}),
admits the following symmetry:

\begin{eqnarray}
\delta \hat{V}({\bf r}_{\perp},z)&=&\delta \hat{V}(-{\bf
r}_{\perp},z)=-\delta
\hat{V}({\bf r}_{\perp},-z)\nonumber\\
\kappa({\bf r}_{\perp},z)&=&\kappa_A\phi_A({\bf r}_{\perp},z)+
\kappa_B\phi_B({\bf r}_{\perp},z)\nonumber\\
&=&\kappa(-{\bf r}_{\perp},z)=\kappa({\bf r}_{\perp},-z) ~,
\end{eqnarray}
where the components of ${\bf r}$ have been written as $({\bf
r}_{\perp},z)$. The self-consistent equations, Eq.~(\ref{maxeq})
-(\ref{inc}), are now solved by a standard procedure of expanding
the functions of position in a complete set of functions with the
above symmetries and those of any specific phase
considered \cite{Matsen94}. We have utilized in our
calculation sets of basis functions containing between 70 and 125
functions, depending upon the value of $\chi N$.

The only parameters entering our calculation are $\chi N$, $f_A$,
$\kappa_A$ and $\kappa_B$, and the rescaled external field $E_{\rm
ext}/{\cal E}$. Comparison of the results with experiment requires
the evaluation of ${\cal E}$ for given $T$ and volume per chain
$v_p$. In addition, the relation between $T$ and $\chi N$ must be
specified.

\section{Results}

As noted earlier, the free energies of lamellar or hexagonal
phases are minimized when the lamellae or the cylinders are
aligned parallel to the electric field because in this orientation
there is no buildup of polarization charge. In the body-centered
cubic (bcc) phase, however, there must be an accumulation of
polarization charge irrespective of the external field direction.
We must determine which field direction produces a phase of
distorted spheres with the lowest free energy, and how large a
field this phase can sustain before a transition to the hexagonal
phase is encountered. It is these issues which we now address.

\subsection{The $R\bar{3}m$ to Hexagonal phase transition}

The symmetry group of the bcc phase is $Im\bar{3}m$ which has
three two-dimensional space {\em subgroups}: $p4mm$ along the
$[100]$ direction, $p6mm$ along the $[111]$ direction, and $p2mm$
along the $[110]$ direction. If the field were applied along
either the $[110]$ or $[100]$ directions the symmetry would be
reduced to $I4/mmm$, while if it were applied along the $[111]$
direction, the symmetry would be reduced to $R\bar{3}m$. The
symmetry in the latter case is of a bcc arrangement of spheres
that has been distorted along the [111] direction. As the
$R\bar{3}m$ group has the $p6mm$ symmetry of the hexagonal phase,
one would suspect that a field applied along the diagonal [111]
direction will result in the lowest free energy. By direct
calculation of these configurations, we find that the $R{\bar 3}m$
phase does indeed have a lower free energy than that of the
$I4/mmm$.

That the electric field favors one orientation of the $Im\bar{3}m$
over another is an effect which is not captured by an expansion of
the free energy to quadratic order in the external field
\cite{Amundson94}. Nonetheless it is instructive to consider the
result of such an expansion. It is obtained by solving the Maxwell
equation $\nabla\cdot[\epsilon_0\kappa(\phi_A,\phi_B){\bf E}]=0$
to second order in ${\bf E}$ to obtain ${\bf E}(\phi_A,\phi_B),$
and evaluating this field from the volume fractions characterizing
the system in the {\em absence} of an external field. The
distortion of the density distribution produced by this field
itself contributes terms to the free energy which are higher order
in $E^2$.
For a phase which is cubic in the absence of an electric field, the
perturbation result can be written as \cite{Amundson93}

\begin{eqnarray}
\label{approx} \frac{F_{\rm pt}({\hat E}_0)}{nk_BT}
& =&-\frac{1}{2}\kappa_0{\hat E}_0^2
\left[1-\frac{1}{12\Omega}\left(\frac{\kappa_A-\kappa_B}{\kappa_0}\right)^2\int
d{\bf r}[\delta\phi_A({\bf r})-\delta\phi_B({\bf r})]^2\right],\nonumber\\
&\equiv& -\frac{1}{2}\kappa_{\rm eff}{\hat E}_0^2, ~
\end{eqnarray}
where $\delta\phi_A=-\delta\phi_B$ is the variation of the local
volume fraction in the zero-field structure, and $\kappa_{\rm
eff}$ is, by definition, the effective dielectric constant for the
structure in the field.

\begin{figure}[ht]
\begin{center}
\epsfxsize=3.5in \epsfbox{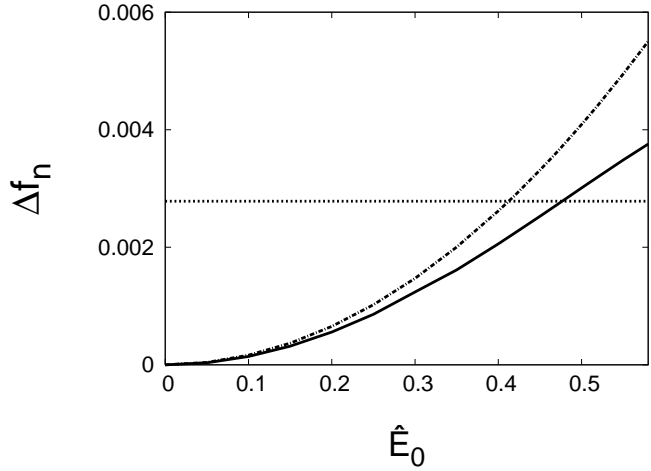}
\end{center}
\caption{The difference, $\Delta f_n$, between the dimensionless
free energy, $f_n$, defined in Eq.~(\ref{fn}) and its value in
zero external field in the bcc phase. It is calculated from the
SCF theory, and is plotted  versus dimensionless external electric
field, ${\hat E}_0=E_0/{\cal E}$, for the hexagonal phase
(horizontal dotted line) and the $R{\bar 3}m$ phase (solid line).
The system is characterized by a $\chi N=15$, $f_A=0.29$. The
dielectric constants are: $\kappa_A=6.0$ (for the PMMA block), and
$\kappa_B=2.5$ (for the PS block), yielding $\kappa_0\simeq 3.52$.
The perturbation theory result for the $R{\bar 3}m$ phase is shown
as a dashed and dotted line. It has a higher free energy.
}
\label{fig:fofphases}%
\end{figure}

We now compare the full SCF solution with this perturbation result. We
choose $\chi N=15$ and $f_A=0.29$, values at which the bcc phase is the
most stable in zero electric field. The dielectric constants are chosen to
make contact with recent experimental systems of polymethylmethacrylate
(PMMA)/polystyrene (PS) diblock copolymer, which is referred to hereafter
as the PMMA-PS system. At experimental temperatures around 160$^\circ$\,C
the dielectric constants appropriate to the PMMA-PS copolymer with PMMA
being the A block and PS the B block are: $\kappa_A=6.0$ (for PMMA), and
$\kappa_B=2.5$ (for PS) \cite{OII00,Xu04,Tsori03} which yield an average
of $\kappa_0\simeq 3.52$.
In Fig.~\ref{fig:fofphases}, we show the difference, $\Delta f_n$,
between the free energy
$f_n(\hat{E}_0)\equiv F_{\rm scf}/nk_BT+\kappa_0{\hat E}_0^2/2$,
Eq.~(\ref{fn}), and its value in zero external field in the bcc
phase. It is shown as a function of $\hat{E}_0$, for the hexagonal
phase and for the $R\bar{3}m$ phase, as calculated from the full
SCF theory and from perturbation theory. The latter is seen to be
adequate for fields smaller than ten to twenty percent of the
natural unit ${\cal E}$ at which ${\hat E}_0=1$. The figure also
shows that there is a transition from the $R\bar{3}m$ to the
hexagonal phase at a value of ${\hat E}_0\simeq 0.477$ as
determined from the full self-consistent calculation. Perturbation
theory underestimates the magnitude of the field needed to bring
about this transition. That the transition is first-order is
easily seen as follows. The average electric and displacement
fields, $E_0$ and $D_0$, are evaluated by taking their spatial
averages over the unit cell. In our case the only non-zero average
components are those in the $z$-direction, and they are related to
the free energy per unit volume according to
\begin{equation} \label{thermo} \frac{\partial F/\Omega}{\partial
E_0}=-D_0 ~,
\end{equation}
or
\begin{equation} \frac{\partial
F/nk_BT}{\partial\hat{E}_0}=-\hat{D}_0 ~.\label{thermo1}
\end{equation}
One sees from
Fig.~\ref{fig:fofphases} that at the phase transition, the free energies
of the $R{\bar 3}m$ and hexagonal phases intersect with different slopes,
therefore the displacement field changes abruptly.

\begin{figure}[ht]
\begin{center}
$\begin{array}{c @{\hspace{0in}}}%
\epsfxsize=2.5in \epsfbox{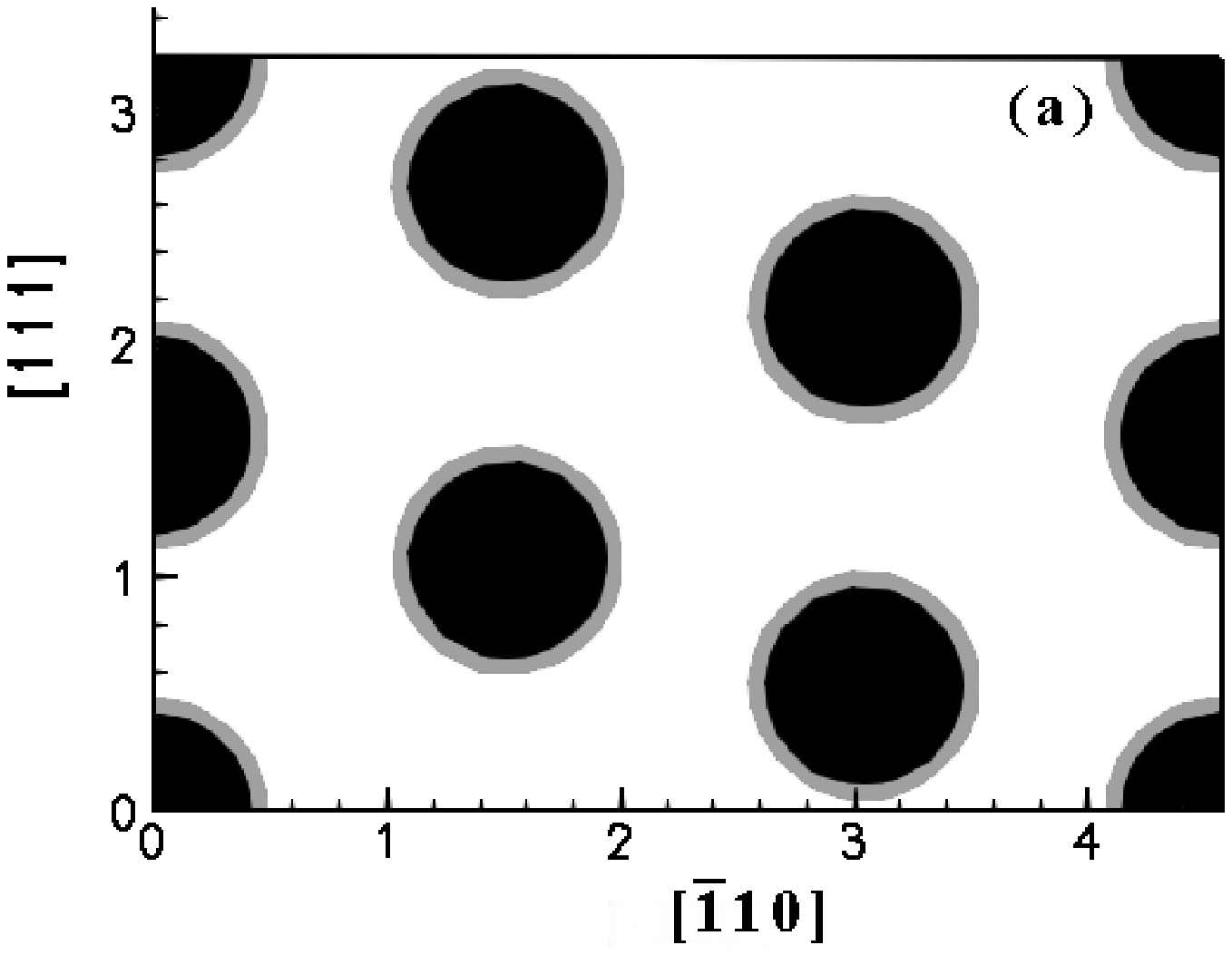}  \\ [0.0in] %
\epsfxsize=2.5in \epsfbox{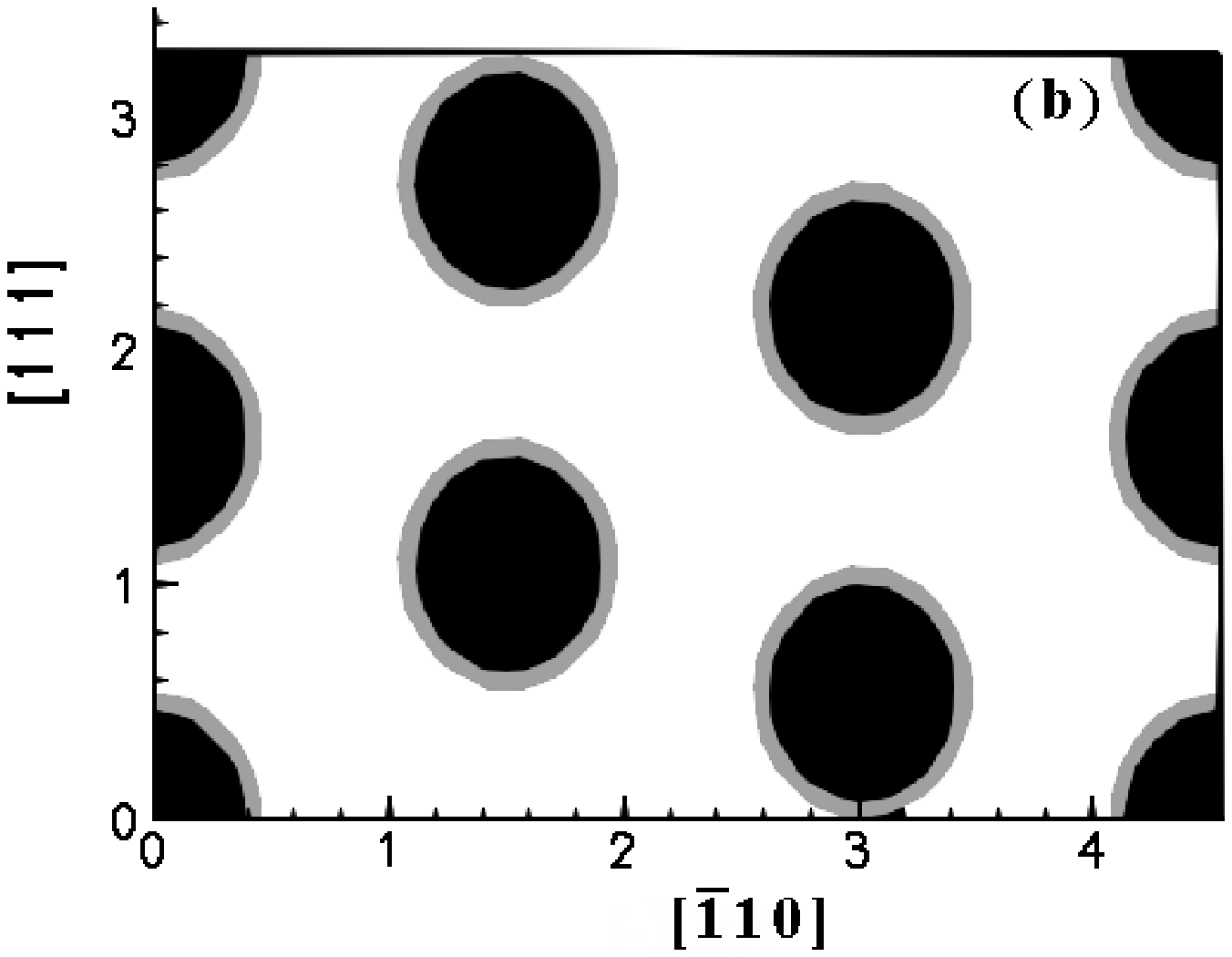}  \\ [0.0in] %
\epsfxsize=2.5in \epsfbox{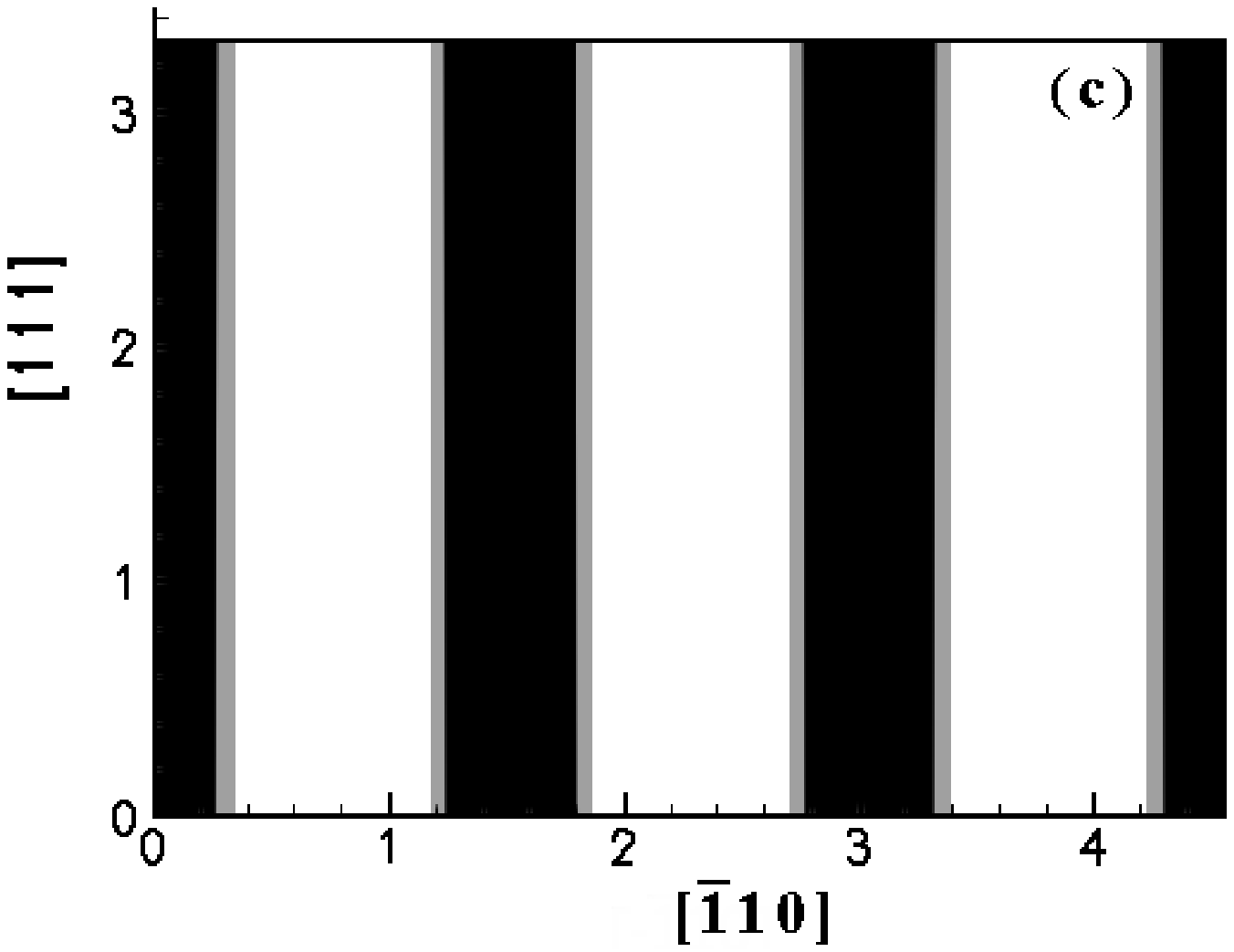}  \\ [0.0in] %
\end{array}$
\end{center}
\caption{ Density profiles for three different phases of a system
characterized by $\chi N=15$ and $f=0.29$, with other parameters as in
Fig.~1.\ (a) the bcc phase  which occurs in zero external field; (b)  the
 $R{\bar 3m}$ phase at
an external electric field $\hat{E}_0=0.470$ just below  the phase
transition to the hexagonal phase which occurs at
$\hat{E}_0=0.477.$   (c) the hexagonal phase, which  is shown for
$\hat{E}_0=0.480$.
The cuts are in the
plane containing the $[111]$ and $[{\bar 1}10]$ directions. In the
black regions, the local volume fractions of component $A$ is
greater than 0.55, in the intermediate regions, it is between 0.55
and 0.45, and in the white  regions, it is less than 0.45. }
\label{fig:profile}%
\end{figure}

As a result of the application of the electric field along the
[111] direction, the spheres of minority component are elongated
in this direction. A density profile of the system in the $R{\bar
3}m$ phase at an external field ${\hat E}_0=0.470$, slightly
smaller than that at the transition to the hexagonal phase, ${\hat
E}_0=0.477$, is shown in Fig.~\ref{fig:profile}(b). At the
transition, the profile changes abruptly to that of the hexagonal
phase, which is also shown in \ref{fig:profile}(c) for ${\hat
E}_0=0.480$. To see the extent of the distortion in the $R{\bar
3}m$ phase, which can be characterized by the aspect ratio of the
distorted spheres, $1.248$, we also present the density profile of
the bcc phase in zero external field, in \ref{fig:profile}(a). The
cuts are in the plane containing the $[111]$ and $[{\bar 1}10]$
directions. %

There are two features of interest that can be seen particularly
clearly from the approximate expression of Eq.~(\ref{approx}). The
first is that in the $R\bar{3}m$ phase, the effective average
dielectric constant, $\kappa_{\rm eff}$, is smaller than
$\kappa_0$. In the hexagonal and disordered phases, however,
$\kappa_{\rm eff}$ is precisely $\kappa_0$.
Therefore, the displacement field
$D_0$ in the $R\bar{3}m$ phase is smaller than in the other two
phases. This is in accord with the change of slope of the free
energy with electric field shown in Fig.~\ref{fig:fofphases} and
Eqs.~(\ref{thermo})-(\ref{thermo1}).

The second concerns the fact that the dielectric constants are
temperature dependent. Therefore, the value of the electric field
needed to bring about a phase transition will also vary with
temperature. The perturbation expression leads one to expect that,
for fields smaller or comparable to ${\cal E}$, the natural
$E$-field scale, the field at the transition will vary as
\begin{equation}
E_{\rm tr}(T)\propto
\frac{[\kappa_0(T)]^{1/2}}{\kappa_A(T)-\kappa_B(T)}
=\frac{[f_A\kappa_A(T)+(1-f_A)\kappa_B(T)]^{1/2}}{\kappa_A(T)-\kappa_B(T)}~.
\end{equation}
%

\subsection{The generalized Claussius-Clapeyron equation}

Before presenting the phase diagram of our A/B block copolymer
system in an $E$-field, we will make use of some general
thermodynamic considerations. In particular, from the differential
of the free energy per unit volume
\begin{equation}
{\rm d}(F/\Omega)=-s{\rm d}T-D_0{\rm d}E_0 ~,
\end{equation}
where $s=S/\Omega$ is the entropy per unit volume, one immediately
derives a Claussius-Clapeyron equation for the slope of the
coexistence line between any two phases
\begin{equation}
\frac {{\rm d}E_0}{{\rm d}T}=-\frac{\Delta s}{\Delta D_0} ~,
\end{equation}
where $\Delta s$ and $\Delta D_0$ are  the differences in
entropies and displacement fields,respectively, in the coexisting
phases. This can be expressed in terms of ${\hat E}_0$, ${\hat
D}_0$ and $\chi N=b/T$ as
\begin{equation}
\label{cc} \frac{{\rm d}{\hat E}_0}{{\rm d}(\chi
N)}=\frac{v_p}{\chi N}\frac{\Delta(s/k_B)}{\Delta{\hat
D}_0}+\frac{{\hat E}_0}{2\chi N} ~.
\end{equation}

\subsection{Phase diagrams}

We now turn to the phase diagram as a function of the inverse
temperature, $\chi N$, the A-monomer fraction, $f_A$, and the
applied external field, $\hat{E}_0=E_0/{\cal E}$. We concentrate
on the portion of the phase diagram involving the phase with
$R{\bar 3}m$ symmetry and the neighboring disordered and hexagonal
phases. In the space of inverse temperature $\chi N$, the fraction
$f_A$, and applied field, the $R{\bar 3}m$ phase occupies a volume
which is bounded by two sheets of first-order transitions: one
from the $R{\bar 3}m$ to the hexagonal phase, the other from the
$R{\bar 3}m$ to the disordered phase. These two sheets of
first-order transitions meet at a line of triple points,
[$\hat{E}_{0,\rm {triple}}(f_A)$, $\chi N_{\rm triple}(f_A)$].
Beyond this line, the $R{\bar 3}m$ phase no longer exists, while
the disordered and hexagonal phases remain. They are separated by
another sheet of first-order transitions which emerges from the
line of triple points. Hence this line is the locus at which all
three sheets of first-order transitions meet.

\begin{figure}[ht]
\begin{center}
\epsfxsize=3.5in \epsfbox{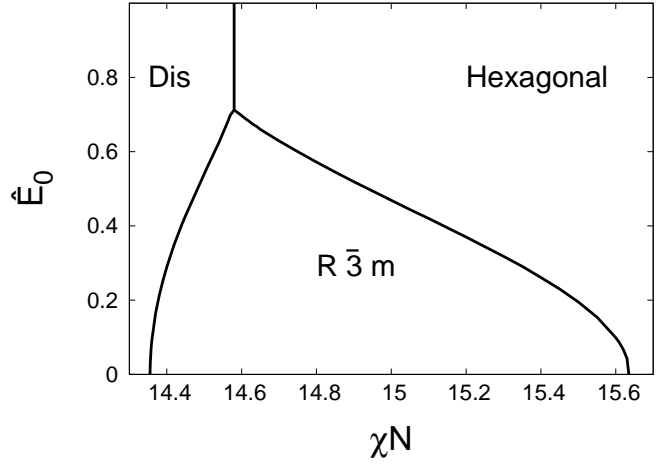}
\end{center}
\caption{Calculated phase diagram of a diblock copolymer with a
volume fraction of $f_A=0.29$ in the presence of an external
electric field.  The phase diagram is shown as a function of the
dimensionless field ${\hat E}_0$ and the interaction parameter
$\chi N$. The triple point is located at $\hat{E}_{0,{\rm
triple}}=0.71$, $\chi N=14.58_{\rm triple}$. Other parameters as
in Fig.~1. }
\label{fig:r3m_constf}%
\end{figure}

In Fig.~\ref{fig:r3m_constf} we show a cut through the phase
diagram at fixed A-monomer fraction, $f_A=0.29$. The cut shows the
phase diagram as a function of the dimensionless electric field
${\hat E}_0$ and $\chi N$. At zero external field, the entropy
difference between the bcc phase and the hexagonal phase is
non-zero, but the difference in displacement field obviously
vanishes. From the Claussius-Clapeyron equation, Eq.~(\ref{cc}),
the slope of the phase boundary between these two phases must be
infinite at zero $E$-field. The same is true for the slope of the
phase boundary between the bcc phase and the disordered phase at
vanishing E-fields. Furthermore, we know from the zero electric
field results that the entropy of the disordered phase is greater
than that of the bcc phase which, in turn, is greater than that of
the hexagonal phase. We also know that the displacement field in
the disordered and in the hexagonal phases is equal to
$\kappa_0\epsilon_0 E_0$.
As we noted earlier, the displacement field in the $R{\bar3}m$
phase is less than this value. This information, together with the
Claussius-Clapeyron, Eq.~(\ref{cc}), implies that the phase
boundary between $R{\bar3}m$ and the disordered phase has a
negative slope, while that between $R{\bar3}m$ and the hexagonal
phase is positive in accord with Fig.~\ref{fig:r3m_constf}.

Moreover, because of the presence of the positive second term in
Eq.~(\ref{cc}), the positive slope of the phase boundary between
disordered and $R{\bar 3}m$ phases will be greater in magnitude,
or steeper, than that between the $R{\bar 3}m$ and hexagonal
phases. This is borne out by Fig.~\ref{fig:r3m_constf}. The three
phases meet at the triple point, above which the phase boundary is
vertical as there is no difference between the displacement fields
of the coexisting disordered and hexagonal phases. The value of
the electric field at the triple point is $\hat{E}_{0,\rm
{triple}}\simeq 0.71$.

To make contact with experiment, we take parameters to fit the
PMMA-PS system  of Ref \cite{UNA00}. With $f_A=0.29$ and a
molecular mass of $3.9\times 10^4$g/mol, and utilizing the known
values of monomeric volumes, we obtain a chain length of $N\simeq
379$
and a volume per PMMA-PS chain of $v_p=61.24$\,nm$^3$ . At
$T=430$\,K this yields ${\cal
E}\equiv(k_BT/\epsilon_0v_p)^{1/2}=104.6$\, V/$\mu$m. Therefore,
the value of the electric field at the triple point is in physical
units $E_{0, {\rm triple}}\approx 74.5$\, V/$\mu$m at this value
of $f_A$ and $T$. One sees from the figure that a transition from
$R\bar{3}m$ to hexagonal phases could be brought about at electric
fields within the interval from this maximum value down to zero,
depending upon the values of $\chi N$ and $f_A$.

The evolution of the phase diagram of Fig.~\ref{fig:r3m_constf}
with A-monomer fraction, $f_A$, is easily understood. As $f_A$
decreases from 0.29, the phase boundary at zero field between
$R{\bar 3}m$ and hexagonal phases moves toward greater values of
$\chi N$ as does the boundary between disordered and $R\bar{3}m$
phases. When $f_A$ is smaller than $f_A^{\rm coex}=0.114$, the
value at which the bcc and hexagonal phases coexist at infinite
$\chi N$ \cite{Matsenwhitmore96}, the boundary between hexagonal
and $R{\bar 3}m$ phases will asymptote with zero slope to an
$f_A$-dependent finite value as $\chi N$ increases without limit.
This zero slope also follows from the Claussius-Clapeyron
equation~(\ref{cc}) due to the fact that the ratio $\Delta
s/\Delta \hat{D_0}$ is finite and $1/\chi
N\rightarrow 0$.%

\begin{figure}[ht]
\begin{center}
\epsfxsize=3.5in \epsfbox{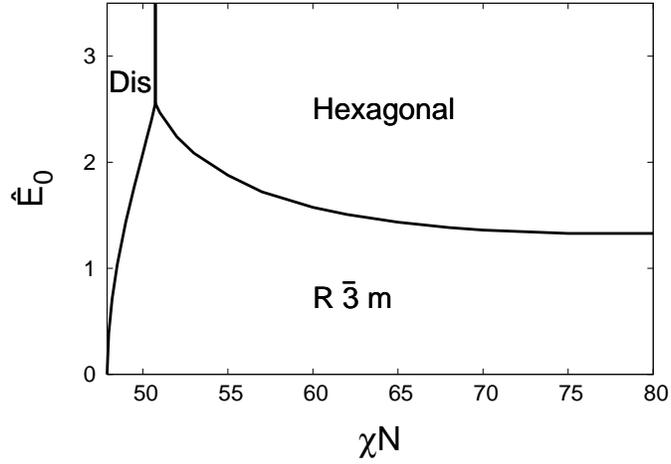}
\end{center}
\caption{Calculated phase diagram of a diblock copolymer in the
presence of an external electric field. Similar to
Fig.~\ref{fig:r3m_constf} but with fraction of the A block,
$f_A=0.1$. The phase diagram is shown as a function of the
dimensionless field ${\hat E}_0$ and the interaction parameter
$\chi N$. The triple point is located at $\hat{E}_{0,{\rm
triple}}=2.56$, $\chi N_{\rm triple}=50.74$.
 }
\label{fig:phasediagram0.1}
\end{figure}

An example of such a phase diagram is shown in
Fig.~\ref{fig:phasediagram0.1}. This figure corresponds to a
system with $f_A=0.1<f_A^{\rm coex}=0.114$, as was investigated
recently in Ref. \cite{Xu04}. In contrast with
Fig.~\ref{fig:r3m_constf}, one sees here that the interval over
which a transition can be observed from $R\bar{3}m$ to hexagonal
phases now extends from the triple point at $\hat{E}_0= 2.56$ down
to a {\em non-zero} minimum value of $\hat{E}_0=1.33$. That is,
for electric fields less than this minimum value, no transition
from the $R{\bar 3}m$ to a hexagonal phase occurs within our
model. For PMMA-PS with $f_A=0.1$, and molecular mass of
$1.51\times10^5$ g/mol, as in Ref.~\cite{Xu04}, we obtain
$N\approx 1458$ and a volume per chain $v_p=239.7$\,nm$^3$.
Therefore at the experimental temperature of $T=430 K$,
the unit of electric field ${\cal E}=52.9$\,V/$\mu$m. In physical
units, then, the triple point occurs at an external field of about
135\,V/$\mu$m and the minimum electric field needed to produce a
transition can be estimated to be 79\,V/$\mu$m.

\begin{figure}[ht]
\begin{center}
\epsfxsize=3.5in \epsfbox{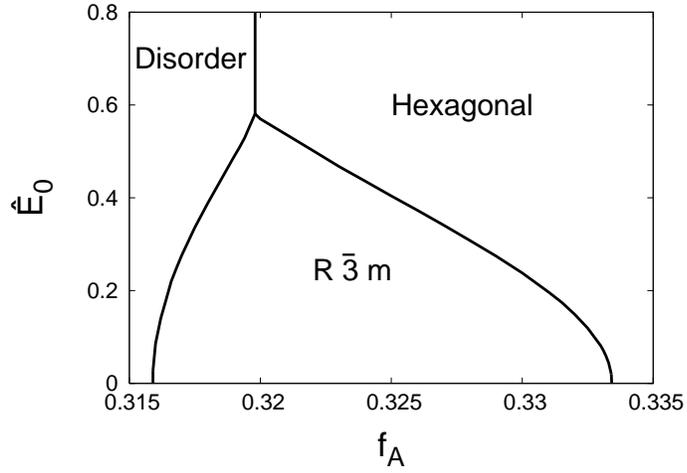}
\end{center}
\caption{Calculated phase diagram for a diblock copolymer as a
function of dimensionless external field and the $A$ mole fraction
parameter, $f_A$. Other parameters are $\chi N=13.3$,
$\kappa_A=6.0$ and $\kappa_B=2.5$. The triple point occurs at
$\hat{E}_{0,{\rm triple}}=0.58$ and $f_{A,{\rm triple}}=0.32$.
 }
 \label{chin13.3}
\end{figure}

In Fig.~\ref{chin13.3} we show a different cut through the phase
diagram in the ($\hat{E}_0$, $f_A$) plane and for a fixed $\chi
N=13.3$. The location of the triple point is $\hat{E}_{0, {\rm
triple}}=0.58$ and $f_{A, {\rm triple}}=0.320$. This figure, and
that of Figs.~\ref{fig:r3m_constf} and \ref{fig:phasediagram0.1},
show that the value of the electric field needed to bring about a
transition from the $R{\bar 3}m$ phase is, for a given $f_A$
fraction, a sensitive function of temperature and, for a given
temperature, a sensitive function of the mole fraction of $A$
block, $f_A$.

\begin{figure}[ht]
\begin{center}
\epsfxsize=3.5in \epsfbox{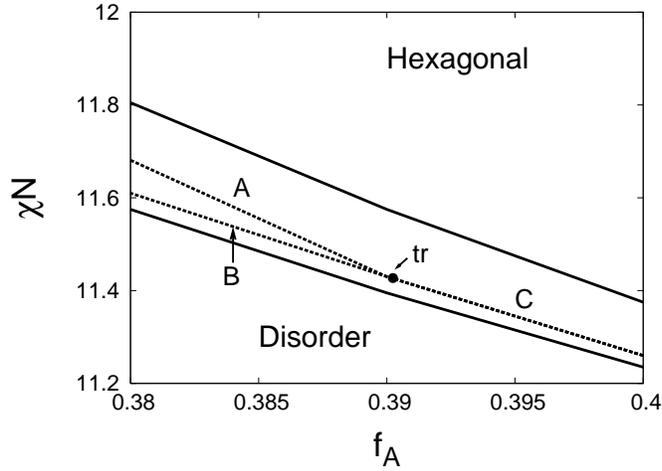}
\end{center}
\caption{Calculated phase diagram at constant electric field,
$\hat{E}_0$. The outer two solid lines are the $E_0=0$
disorder-to-bcc and bcc-to-hexagonal phase boundaries. Between
them we show three other transition lines for  ${\hat E}_0=0.2$.
They are the
$R\bar{3}m$-to-Hexagonal, ($A$),  $R\bar{3}m$-to-Disorder ($B$),
and Disorder-to-Hexagonal
transition ($C$). These three lines meet at $tr$, the triple
point: $f_{A. {\rm triple}}=0.390$ and $\chi N_{\rm
triple}=11.43$.
} \label{fig:r3m_constE}
\end{figure}

Figure~\ref{fig:r3m_constE} shows two cuts through the phase
diagram at constant electric field, ${\hat E}_0=0$, and ${\hat
E}_0=0.2$. The solid line at lower $\chi N$ shows the phase
boundary at zero-field between disordered and bcc phases while the
solid line at larger $\chi N$ shows the zero-field phase boundary,
between the bcc and hexagonal phases. The dashed lines between
them show the phase boundaries for ${\hat E}_0=0.2$. The line denoted $B$
is the boundary between the disordered and the $R{\bar 3}m$ phase
of distorted spheres, $A$ is the boundary between $R{\bar 3}m$ and
hexagonal phases. These boundaries meet at the triple point, $tr$,
which occurs at $\chi N\approx 11.43$, and $f_A\approx 0.39.$ For
larger values of $f_A$, there is a line, $C$, of transitions
directly
from the disordered to the hexagonal phase. As the external field
increases still further, the triple point recedes to larger values
of $\chi N$ leaving behind only the line of direct transitions
between disordered and hexagonal phases. Note that, except for the
location of its terminus at the triple point, this boundary is
independent of the applied field as it contributes to the free
energy of both of these phases equally. The dielectric
constants used to generate this figure are the same as those used
in previous figures.

For completeness, we have also examined the case in which the
dielectric constants of the minority and majority components are
interchanged as compared with Fig.~\ref{fig:r3m_constf}. Namely,
the majority component with, $f_A=0.71$ has the larger dielectric
constant of $\kappa_A=6.0$ and the minority the smaller value of
$\kappa_B=2.5$. We find that the $R{\bar3}m$ phase is now somewhat
more stable with respect to the hexagonal phase, so that the value
of the external electric field needed to bring about a transition
from the former to the latter phase is increased. We note that
this interchange increases the average value of the dielectric
constant, so that all phases have a lower free energy due to the
factor of $-\kappa_0\hat{E}_0^2/2$ which it contains. However, it
is not {\em a priori} obvious that the $R{\bar 3}m$ phase would
have its free energy lowered by more than that of the hexagonal
phase by this interchange. In addition, we have determined that
the spheres of minority component and lower dielectric constant
distort in the [111] direction just as in the case when the
minority component has the larger dielectric constant. The above
effects are not captured by the perturbation result of
Eq.~(\ref{approx}) which is invariant under the interchange of
$\kappa_A$ and $\kappa_B$.

\section{Concluding remarks}

In sum, we have calculated the phase diagram of a block copolymer
system in an external electric field which couples to the diblocks
through the difference in their dielectric constants. We have
employed a fully self-consistent field approach in which the
relevant Maxwell equation is treated on an equal footing with the
other self-consistent equations. We have determined that the
body-centered cubic phase will preferentially align along the
[111] direction causing its symmetry to be reduced to $R{\bar
3}m$. The electric field can induce phase transitions between this
phase and either the disordered or the hexagonal phase. The
strength of the field needed to induce such transitions is a
sensitive function of the parameters of the system, such as its
temperature and its chain architecture, which in the case of
linear diblocks is quantified simply by the mole fraction, $f_A$.

For parameters that fit the experimental PMMA-PS diblock copolymer
system investigated recently \cite{Xu04}: $f_A=0.1$,
$v_p=239.7$\,nm$^3$, and $T=430$\,K, we find that an electric
field of at least $70-80$\, V/$\mu$m would be needed to observe a
transition to the hexagonal phase. This contrasts with the
reported existence of such a phase transition under an applied
field of only 40\,V/$\mu$m. There are several possible
explanations of the difference between the experimental results
and the theoretical ones presented here.

Our model employs a linear constitutive relation between
dielectric constant and volume fractions, and characterizes the
PMMA-PS system by a few general parameters, the PMMA mole fraction
$f_A$, and the interaction parameter $\chi N$. It further assumes
equal volumes for both monomers and equal Kuhn lengths for them.
One knows that deviations from the last assumption certainly shift
the locations of the phase boundaries \cite{Matsen94b}. It is
plausible that at rather asymmetric volume fractions of $f_A=0.1$,
the model provides only semi-quantitative agreement with the
experimental PMMA-PS phase diagram.  Any difference in the
theoretical and experimental phase diagrams at zero electric field
will, in the presence of a non-zero one, manifest itself in a
difference in relative stability of the various phases. Given the
sensitive dependence on the phase boundaries of the minimum
external field needed to bring about a phase transition,
differences between the general theory and the experimental result
are to be expected. At present, the phase diagram of the PMMA-PS
system of Ref.~\cite{Xu04} is not yet known. When additional
experimental information becomes available, one will also need to
determine the relationship between the temperature and the $\chi
N$ interaction parameter in order to convert the phase diagram
calculated here to practical units so that it can be
compared directly to the experimental one.

Lastly, we have employed a simple coupling between the system and
the external field via the difference in dielectric constants of
the copolymer blocks. Other couplings are possible
\cite{Gurovich94,Onuki95}.  Just such an additional coupling, to
mobile ions, has been suggested by Tsori et al. \cite {Tsori03}
and is discussed also in Ref. \cite{Xu04}. A minute fraction of
mobile ions embedded in the minority PMMA fraction and not in the
majority PS can lead to an enhanced response of the PMMA-PS system
to external electric fields with moderate magnitude. It could also
change the phase diagram quantitatively resulting in a substantial
lowering of the triple-point value of the electric field.
Additional experiments, particularly on copolymers with the same
PMMA-PS blocks, but at different temperatures or values of the
architectural parameter $f_A$, would be most useful to shed
additional light on the comparison of theory and experiment.
In particular, a comparison of the two PMMA-PS systems of
Ref.~\cite{UNA00} and Ref.~ \cite{Xu04} would be enlightening because,
as Figs.~\ref{fig:r3m_constf} and ~\ref{fig:phasediagram0.1} show, they
are predicted here to exhibit significantly different phase behavior in
an external electric field.
\section*{Acknowledgment}

We are indebted to Ludwik Leibler, Tom Russell, Yoav Tsori and
Ting Xu for illuminating exchanges. Partial support from the
National Science Foundation under grant No. 0140500, the National
Science Foundation IGERT fellowship from the University of
Washington Center of Nanotechnology, the United States-Israel
Binational Science Foundation (BSF) under grant No. 287/02 and the
Israel Science Foundation under grant No. 210/01 is gratefully
acknowledged.


\clearpage

\clearpage

\clearpage

\clearpage

\clearpage

\clearpage

\end{document}